\documentclass[twoside,twocolumn,tightenlines,superscriptaddress, showpacs,aps,prb]{revtex4}
\usepackage{graphicx}
\usepackage{wasysym}
\usepackage{amssymb}
\usepackage[makeroom]{cancel}
\newcommand{\be}{\begin{equation}}
\newcommand{\ee}{\end{equation}}
\usepackage[unicode=true,pdfusetitle,
 bookmarks=true,bookmarksnumbered=false,bookmarksopen=false,
 breaklinks=false,pdfborder={0 0 1},backref=false,colorlinks=false]
 {hyperref}
\hypersetup{
 colorlinks,linkcolor=blue,citecolor=blue,urlcolor=blue}

\begin{document}

\preprint{APS/123-QED}

\title{Spin Hall angle fluctuations in a disorder device}

\author{F. A. F. Santana}
\affiliation{Departamento de F\'{\i}sica, Universidade Federal Rural de Pernambuco, 52171-900, Recife, PE, Brazil}

\author{J. M. da Silva}
\affiliation{Departamento de F\'{\i}sica, Universidade Federal Rural de Pernambuco, 52171-900, Recife, PE, Brazil}

\author{T. C. Vasconcelos}
\affiliation{Departamento de F\'{\i}sica, Universidade Federal da Para\'iba, 58297-000 Jo\~ao Pessoa, Para\'iba, Brazil}

\author{J. G. G. S. Ramos}
\affiliation{Departamento de F\'{\i}sica, Universidade Federal da Para\'iba, 58297-000 Jo\~ao Pessoa, Para\'iba, Brazil}

\author{A. L. R. Barbosa}
\email{anderson.barbosa@ufrpe.br}
\affiliation{Departamento de F\'{\i}sica, Universidade Federal Rural de Pernambuco, 52171-900, Recife, PE, Brazil}

\date{\today}

\begin{abstract}
We investigate a disorderly mesoscopic device that supports spin-orbit interaction. {The system is connected to four semi-infinite leads embedded in the Landauer-Buttiker setup for quantum transport} and, according to our analysis, exhibits spin Hall angle fluctuations. We show analytically and numerically the fingerprint of the universal fluctuation of the polarization mediated by the conversion of charge current into spin current. Our investigation shows the complete compatibility of our analytical and numerical results with the most recent experiments. {Furthermore, we show nonzero and universal features of spin Hall effect in Rashba 2DEG with disorder.} All the results show the relevance of microscopic parameters for electronic transport with charge-spin conversion and, in many cases, inevitably lead to universal numbers.
\end{abstract}


\pacs{Valid PACS appear here}
\maketitle


{\it Introduction} - The spin-orbit interaction (SOI) is a relativistic effect which is found in many branches of condensed matter physics [\onlinecite{101038nature19820,RevModPhys.89.025006,10.1038/nmat4360,Vignale,Nikolic2009}]. Such coupling permeates the history of quantum mechanics through its numerous manifestations and applications that include the hyperfine structure in atomic spectroscopy, the modification of shell models in nuclear physics and, more recently, spintronics [\onlinecite{Nikolic2005,Nikolic2007,Milletari,Liu555,PhysRevLett.124.087702,Wang_2018,PhysRevLett.116.196602,PhysRevB.97.245414,doi:10.1063/1.5010973,PhysRevB.93.115120}]. One of the most relevant manifestations of the spintronic is the spin Hall effect (SHE) [\onlinecite{RevModPhys.87.1213,Niimi_2015}], which was proposed in the Refs. [\onlinecite{DYAKONOV1971459,PhysRevLett.83.1834}] and measured for the first time in the Refs. [\onlinecite{Kato1910,PhysRevLett.94.047204}].
The main mechanism underlying the effect is an electric field applied to the device in the longitudinal direction generating a pure longitudinal charge current, as usual. However, the up spins electrons are deflected to a diametrically opposite side of the down spins electrons, in the same amount, giving rise to a pure transversal spin Hall current due to SOI. To quantify the efficiency of charge-to-spin conversion, it is commonly used the spin Hall angle (SHA), which is defined as ratio between the vertical spin Hall current and longitudinal charge current. Its experimental values can range between 0.01$\%$ to 58$\%$ for different materials in disorderly regime [\onlinecite{10.1038/ncomms1640,PhysRevB.87.224401,PhysRevB.94.060412,doi:10.1021/acs.nanolett.6b02334,PhysRevB.98.094433,doi:10.1063/1.4753947,PhysRevB.101.094435,PhysRevLett.122.217701,PhysRevApplied.10.031001,PhysRevLett.112.197201,PhysRevB.96.060408,doi:10.1063/1.5079813,10.1038/nphys3304}].

The SHE fluctuations were theoretically investigated in the Refs. [\onlinecite{PhysRevLett.97.066603}] in a disordered four-leads
device using a tight-binding model. The authors showed the presence of a universal spin Hall conductance fluctuations with a universal number $\textbf{rms}[G_{\text{sH}}] = 0.18e/4\pi$ in the presence of the SOI. Motivated by this numerical result, the authors of Ref. [\onlinecite{PhysRevLett.98.196601}] were able to recover this universal number analytically using the Ladauer-Buttiker formulation (LBF) [\onlinecite{PhysRevB.86.155118}] and the random matrix theory (RMT) [\onlinecite{Meh2004}]. Furthermore, they demonstrated the universal behavior established with the circular symplectic ensemble (CSE) in the framework of RMT. In the current literature, there are many SHA theoretical studies [\onlinecite{PhysRevLett.116.196602,PhysRevB.91.045407}], however a theoretical investigation of SHA fluctuations concatenating both by numerical calculation and  all the analytical results are completely missing.    

Given this scenario, a relevant question that remains open is: what information regarding electronic transport is provided by a measurement of SHA fluctuations? We will show, analytically using LBF, RMT, 
DMPK equation [\onlinecite{Mello}] and central limit theorem (CLT) [\onlinecite{Reichl:101976}] that the SHA deviation is a function of only three variables in the disorderly regime with strong SOI: the sample thickness, longitudinal length and the free electron path. In addition to these results, we show that if the sample length is long enough, the SHA maximum deviation holds a universal relation with dimensionless conductivity $\Theta_{\text{sH}} \times \sigma = 0.18$ which is independent of the material and its specific features. This universal relation is supported by five different experimental data and a numerical calculation. {Furthermore, despite the consensus of a vanishing SHE due to disorder [\onlinecite{Vignale,Milletari}], we show that the zero SHE are irrelevant for realistic finite-size systems where self-averaging over an infinite system size is avoided [\onlinecite{Nikolic2009,Nikolic2007}].}

\begin{figure}
\includegraphics[scale = 0.35]{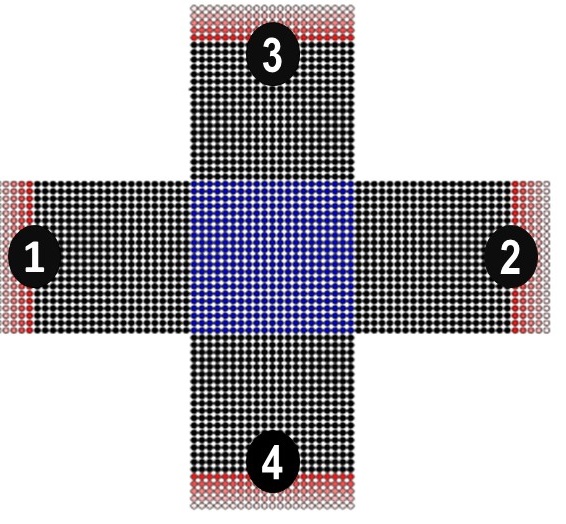}
\caption{The spin Hall device design. The scattering sample with disorder and strong SOI (blue) is connected to {four semi-infinite leads}.}\label{sample}
\end{figure}

{\it SHA fluctuations} -  The device is designed with four semi-infinite leads (black) connect to a scattering region with disorder and strong SOI (blue) as depicted in the Fig.(\ref{sample}). An electric potential difference $V$ is applied between the leads $1$ and $2$, which gives rise to a pure longitudinal charge current. 

From the LBF, the Refs. [\onlinecite{Nikolic2005,Nikolic2007,PhysRevLett.98.196601}] were able to obtain the following expression for the vertical spin Hall current
\begin{eqnarray}
I^{s}_{i,\alpha} = \frac{e^2}{h}\left[\left(\tau^\alpha_{i2}-\tau^\alpha_{i1}\right)\frac{V}{2}
- \tau^\alpha_{i3}V_3 + \tau^\alpha_{i4}V_4\right],\; i={3,4},
\label{Is}
\end{eqnarray}
and also for longitudinal charge current
\begin{eqnarray}
I^c &=& \frac{e^2}{h}\left[\left(4N+\tau^0_{12}+\tau^0_{21}-\tau^0_{11}-\tau^0_{22}\right)\frac{V}{4}\right.\nonumber\\
&+& \left.\left(\tau^0_{23}-\tau^0_{13}\right)\frac{V_3}{2} + \left(\tau^0_{24}-\tau^0_{14}\right)\frac{V_4}{2}\right].
\label{Ic}
\end{eqnarray}
The dimensionless integer $N$ is the number of propagating wave modes in the leads, which is proportional to both the lead width ($W$) and the Fermi vector ($k_F$) through the equation $N = k_F W/\pi$, while $V_{3,4}$ are the vertical leads potential. The transmission coefficients $\tau^\alpha_{ij}$ can be obtained from  transmissions and reflections blocks of the corresponding device scattering  $\mathcal{S}$-matrix as
$$
\tau_{ij}^{\alpha} =\textbf{Tr}\left[\left(\mathcal{S}_{ij}\right)^{\dagger}\sigma^\alpha\mathcal{S}_{ij}\right], \quad 
\mathcal{S}=
\left[\begin{array}{cccc}
r_{11}&t_{12}&t_{13} &t_{14}\\
t_{21}&  r_{22}&t_{23}&t_{24}\\
t_{31}&  t_{32}&r_{33}&t_{34}\\
t_{41}&  t_{42}&t_{43}&r_{44}
\end{array}\right],
$$
with $\sigma^0$ and $\sigma^\alpha$ denoting the identity and Pauli matrices, respectively, with polarization direction $\alpha = {x,y,z}$.

The SHA is defined as the ratio between vertical spin Hall and longitudinal charge currents
\begin{eqnarray}
\Theta_{\text{sH}}=\frac{I^{s}}{I^c}.\label{I/I}
\label{angle}
\end{eqnarray}
To develop the ensemble average of the Eq.(\ref{I/I}), the CLT can be implemented. Hence, taking high Fermi energy limit $E\gg0$, which means that device thickness is large $N\gg1$, the Eq.(\ref{angle}) can be expanded as
\begin{equation}
\left\langle\Theta_{\text{sH}}\right\rangle = \frac{\langle{I^s}\rangle}
{\langle{I^c}\rangle}+
\frac{\langle{{\delta}I^s}\rangle{\langle I^c\rangle}-
\langle{{\delta}I^c}\rangle{\langle I^s}\rangle}
{\langle{I^c}\rangle^2}+\mathcal{O}(N^{-1})\label{exp}
\end{equation}
This methodology is often used to electronic transport in RMT [\onlinecite{RevModPhys.69.731,PhysRevB.88.245133,PhysRevE.90.042915,PhysRevB.84.035453,Barbosa_2010,PhysRevB.86.235112}]. As showed in the Refs.[\onlinecite{PhysRevLett.98.196601,PhysRevB.86.235112}] the spin Hall current average is null, $\langle{I^s}\rangle = \langle{{\delta}I^s}\rangle = 0$, which leads us to deduce
\begin{equation}
\left\langle\Theta_{\text{sH}}\right\rangle = 0.\label{TM}
\end{equation}
The Eq.(\ref{TM}) for the SHA implies a Gaussian distribution with maximum in zero and also that all relevant information may be contained in its fluctuations. The device under study is disorderly, which induces universal spin Hall and charge currents fluctuations [\onlinecite{PhysRevLett.97.066603}]. Hence, it is reasonable to be expected that the SHA has universal fluctuations. In the usual way, we define the SHA deviation as
$$
\textbf{rms}[\Theta_{\text{sH}}]=\sqrt{\left\langle\Theta_{\text{sH}}^2\right\rangle-\left\langle\Theta_{\text{sH}}\right\rangle^2} = \sqrt{\left\langle\Theta_{\text{sH}}^2\right\rangle},\nonumber
$$
We follow the same methodology above to the ensemble average and obtain
\begin{eqnarray}
\left\langle\Theta_{\text{sH}}^2\right\rangle &=& \frac{\langle{I^s}\rangle^2}
{\langle{I^c}\rangle^2}+2\frac{\langle{{\delta}I^s}\rangle\langle{I^s}\rangle{\langle{I^c}\rangle}-
\langle{{\delta}I^c}\rangle\langle{I^s}\rangle^2}
{\langle{I^c}\rangle^3}\nonumber\\&+&
\frac{\langle{{\delta}{I^s}^2}\rangle\langle{I^c}\rangle^2 +\langle{{\delta}{I^c}^2}\rangle{\langle{I^s}\rangle}^2-2\langle{{\delta}I^s{\delta}I^c}\rangle{\langle{I^s}\rangle}\langle{I^c}\rangle}
{\langle{I^c}\rangle^4}\nonumber\\
&+&\mathcal{O}(N^{-3}).\nonumber
\end{eqnarray}
Using the zero mean again for the current, $\langle{I^s}\rangle = 0$, it simplifies to 
\begin{eqnarray}
\textbf{rms}[\Theta_{\text{sH}}] &=& \sqrt{\frac{\langle{{{\delta}{I^s}^2}}\rangle}
{\langle{I^c}\rangle^2}}, \label{Tm}
\end{eqnarray}
that is, we can infer the SHA deviation with the knowledge the spin Hall current fluctuations and the charge current average. 

Applying the diagrammatic method [\onlinecite{doi:10.1063/1.531667}] to scattering matrices in the circular symplectic ensemble (strong SOI), it was obtained for spin Hall current fluctuation the expression [\onlinecite{PhysRevLett.98.196601,PhysRevB.86.235112}]
\begin{eqnarray}
\langle{{{\delta}{I^s}^2}}\rangle = \left(\frac{e^2V}{h}\right)^2\left[\frac{1}{32}+\mathcal{O}(N^{-1})\right].\label{Is2}
\end{eqnarray}
At this point, we must invoke calculations that incorporate length scales that are not covered by diagrammatic method [\onlinecite{doi:10.1063/1.531667}]. The longitudinal charge current average is appropriately described by the result provided by DMPK [\onlinecite{Mello,RevModPhys.69.731}]
\begin{eqnarray}
\langle{{{I^c}}}\rangle = \frac{e^2V}{h}\left[\frac{N}{1+\frac{L}{l_e}}+\mathcal{O}(N^{-1})\right],\label{Ic2}
\end{eqnarray}
where $L$ and $l_e$ are device longitudinal length and free electron path, respectively. The limit $L/l_e \gg 1$ leads to diffusive regime while $L/l_e \ll 1$ ballistic regime assuming that phase coherence length $L_\phi$ satisfies $L_\phi>L$.
Substituting the Eqs.(\ref{Is2}) and (\ref{Ic2}) in the Eq.(\ref{Tm}), we obtain
\begin{eqnarray}
\textbf{rms}[\Theta_{\text{sH}}] = \frac{0.18}{N} \left(1+\frac{L}{l_e}\right)\label{rmsT}.
\end{eqnarray}
The Eq.(\ref{rmsT}) is the main outcome of this work, which expresses the universal fluctuation as a function of three variables relevant to the electronic transport. The Eq.(\ref{rmsT}) drives to two important interpretations: 1) disorder increase the SHA, the more scattering the spin carrier suffers the greater the charge-spin conversion; 2) decreasing of device thickness $N$ increasing SHA. The authors of Ref.[\onlinecite{PhysRevB.91.045407}] have used the Drude model and found that the SHA can be enhanced by decreasing film thickness, which is in accordance with Eq.(\ref{rmsT}).

Taking the limit $L/l_e\ll1$, the SHA attains a maximum deviation with the limit of Eq.(\ref{rmsT}) resulting in
\begin{eqnarray}
\Theta_{\text{sH}} \times g = 0.18, 
\end{eqnarray}
which is valid to chaotic ballistic billiard and accordingly $g=N$ is the dimensionless conductance. Furthermore, taking the limit $L/l_e\gg1$, the Eq.(\ref{rmsT}) can be written as a function of dimensionless conductivity $\sigma =Nl_e/L$ as 
\begin{eqnarray}
\Theta_{\text{sH}} \times \sigma = 0.18, \label{Tc}
\end{eqnarray}
which indicates the decrease in SHA as a power law as a function of conductivity for films with strong SOI in disorderly regime. Moreover, the Eq.(\ref{Tc}) means that the product between $\Theta_{\text{sH}} $ and $\sigma$ has a universal value 0.18, which is independent of the material and its specific features. 

\begin{figure}
\includegraphics[scale = 0.23]{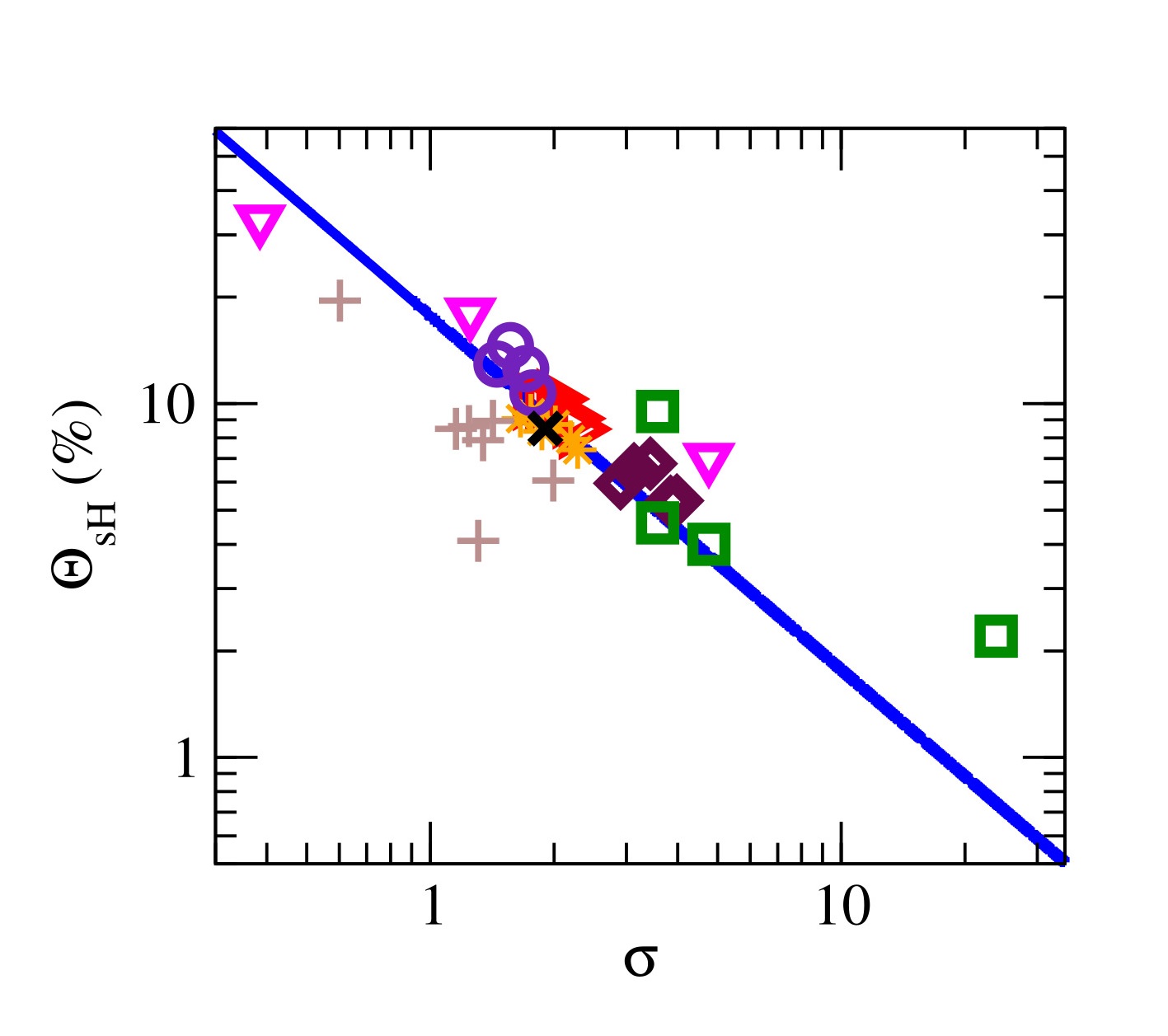}
\caption{The figure shows the SHA $\Theta_{sH} (\%)$ as a function of dimensionless conductivity $\sigma$. The symbols circle, star, diamond and triangle right are experimental data  obtained from the Ref.[\onlinecite{PhysRevB.94.060412}]. The experimental square, plus and triangle down and times symbols are obtained from Refs.[\onlinecite{doi:10.1021/acs.nanolett.6b02334,PhysRevB.98.094433,doi:10.1063/1.4753947,PhysRevB.101.094435}], respectively. The continuum line (blue) is the analytical result of the Eq.(\ref{Tc}).}\label{expprb}
\end{figure}

{\it Experimental analysis} - The Fig.(\ref{expprb}) shows $\Theta_{\text{sH}} (\%)$ as a function of dimensionless conductivity $\sigma$. The symbols circle, star, diamond and triangle right are experimental data obtained from the Fig.(4) of the Ref.[\onlinecite{PhysRevB.94.060412}]. Pt films were used in the moderately dirt regime. The conductivity axis of experiment was normalized as $\sigma=\sigma_{\text{exp}}(\Omega^{-1}\cdot \text{cm}^{-1})/10^{4}(\Omega^{-1}\cdot \text{cm}^{-1})$. 

The experimental square symbols are obtained from the Table 1 of the Ref.[\onlinecite{doi:10.1021/acs.nanolett.6b02334}] for films of NiFe/Pt, CoFe/Pt CoFe/Pd and CoFe/Au from $\rho_N(\mu\Omega\cdot\text{cm})$ and $\Theta_{\text{SHE}}$ (1D-anlylical) ($\%$) columns.  The plus symbols are obtained from the Fig.(2.a,b) of the Ref.[\onlinecite{PhysRevB.98.094433}] for films based on W by mixing with Hf with concentration $\geq 0.7$. Moreover, the triangle down symbols are obtained from Table 1 of the Ref.[\onlinecite{doi:10.1063/1.4753947}] for $\beta$-W thin films, while the times symbol are obtained from the Ref.[\onlinecite{PhysRevB.101.094435}] for $p$-Si thin film.

In the same Fig.(\ref{expprb}), we plot the Eq.(\ref{Tc}) as a continuum line (blue) and, as depicted, we conclude the compatibility between the five experiments [\onlinecite{PhysRevB.94.060412,doi:10.1021/acs.nanolett.6b02334,PhysRevB.98.094433,doi:10.1063/1.4753947,PhysRevB.101.094435}] and our analytical results   follow satisfactorily the universal relation $\Theta_{\text{sH}} \times \sigma = 0.18$.

\begin{figure}
\includegraphics[scale = 0.35]{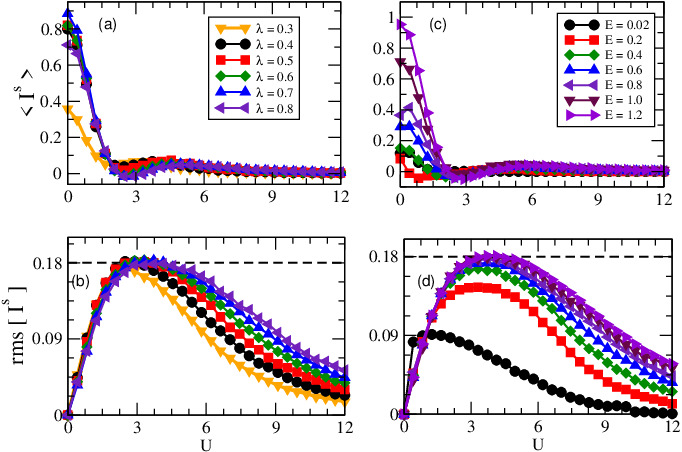}
\caption{The figures (a,c) show the spin current average while (b,d) show the spin current deviation as a function of the disorder $U$. The figures (a,b) are for different values of SOI $\lambda$ at fix $E=1$, while (c,d) are for different values of $E$ at fix $\lambda=0.8$. In both cases the spin Hall current deviation results in $\text{rms}[I^s] = 0.18$ (dashed line), Ref.[\onlinecite{PhysRevLett.97.066603}].}\label{figIs}
\end{figure}

{\it Numerical results} - We developed a numerical calculation of SHA fluctuations and we established a direct comparison with the Eqs.(\ref{TM}) and (\ref{rmsT}). The device design is depicted in the Fig.(\ref{sample}) and the tight-binding Hamiltonian of scattering region (blue) is [\onlinecite{PhysRevB.74.085327,PhysRevB.98.155407}]
\begin{eqnarray}
 H&=&-t\sum_{\langle i,j\rangle,\sigma}c_{i\sigma}^{\dagger} c_{j\sigma} + \sum_{i\sigma} \left(4t + \epsilon_i\right) c_{i\sigma}^\dagger c_{i\sigma}\nonumber\\
 &-& i \lambda \sum_{\langle i,j \rangle}\left(c_{i}^{\dagger}\sigma_y c_{j}-c_{i}^{\dagger}\sigma_x c_{j}\right) \label{TBH}
\end{eqnarray}
The fist term represents the usual nearest-neighbor interaction, where $c_i$ ($c_i^\dagger$) are the annihilation (creation) operators and $t=\hbar^2/2m^* a^2$ is the nearest-neighbor hopping energy [\onlinecite{numero}]. The second one is an Anderson disorder term. The disorder is realized by an electrostatic potential $\epsilon_i$ which varies randomly from site to site according to a uniform distribution in the interval $\left(-U/2,U/2\right)$, where $U$ is the disorder strength. The last one, $\lambda=\hbar \alpha_R/2a$ describes the strength of the Rashba SOI. The numerical calculations [\onlinecite{numero}] implemented in the KWANT software [\onlinecite{Groth2014}]. 

\begin{figure}
\includegraphics[scale = 0.35]{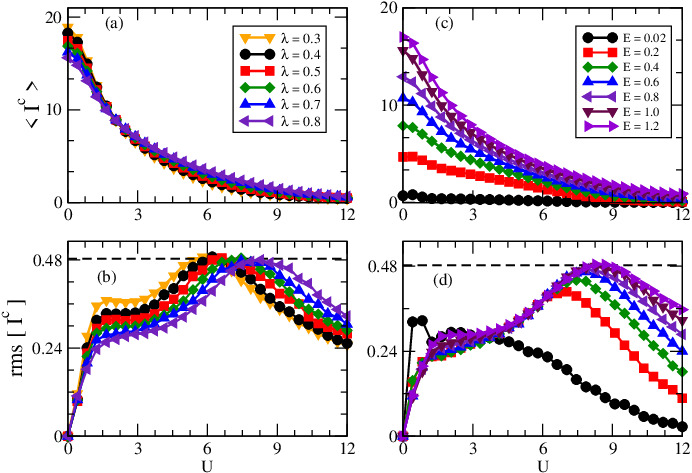}
\caption{The figures (a,c) show the charge current average while (b,d) show the charge current deviation in function of disorder $U$. Figures (a,b) are for different SOI values $\lambda$ at fix $E=1$, while (c,d) are for different values of $E$ at fix $\lambda=0.8$. In both cases the charge current deviation hold a maximum in $\text{rms}[I^{c}] = 0.48$ (dashed line). }\label{figIc}
\end{figure}

The Fig.(\ref{figIs}) shows the universal spin Hall current fluctuation in agreement with the previous numerical [\onlinecite{PhysRevLett.97.066603}] and analytical [\onlinecite{PhysRevLett.98.196601}] results. The Figs.(\ref{figIs}.a,c) represents the spin Hall current average, Eq.(\ref{Is}), as a function of disorder $U$ for different values of $\lambda$ and energy, respectively. In both cases, we observe oscillations in the tails of the spin Hall current average, which were not announced before. The oscillations have as the underlying mechanism the fluctuations in potentials $V_{3,4}$. Furthermore, the Figs.(\ref{figIs}.b,d)
show the spin Hall current deviation as a function of $U$. In the former, the energy was fixed in $E=1$ for different SOI values $\lambda$; In all cases, the maximum deviations are $\textbf{rms}[I^s] = e^2V/h \times 0.18$ (dashed line), as expected. In the latter, the SOI value was fixed in $\lambda = 0.8$ for different energy values; For low energy ($E=0.02$) the spin Hall current has its minimum deviation, while for high energies ($E \geq 0.6$) it has its maximum deviation.

\begin{figure}
\includegraphics[scale = 0.35]{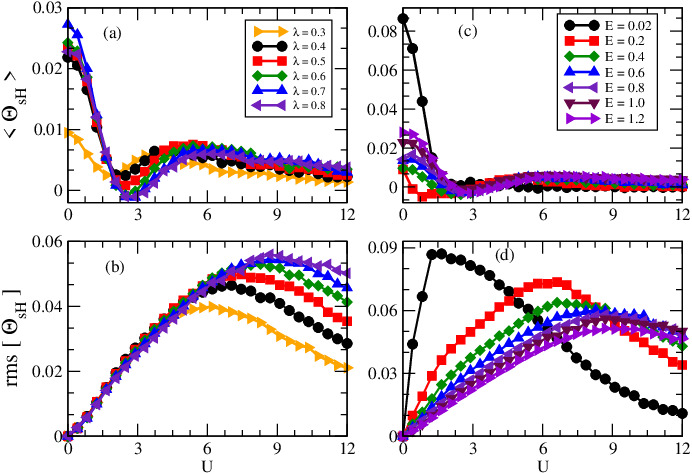}
\caption{The figures (a,c) show the SHA average while (b,d) show the one deviation in function of disorder $U$. Figures (a,b), each curve is for a different value of SOI $\lambda$ at fix energy $E=1$.  Figures (c,d), each curve is for a different value of $E$ at fix $\lambda=0.8$. }\label{figASH}
\end{figure}

The longitudinal charge current behavior, Eq.(\ref{Ic}), is depicted in the Fig.(\ref{figIc}). The Figs.(\ref{figIc}.a,c) show the charge current average as a function of $U$ for different values of $\lambda$ and energy, respectively, while the Figs.(\ref{figIc}.b,d) are their respective deviations. Differently of the spin Hall current average, depicted in the Figs.(\ref{figIs}.a,c), the charge current average does not present oscillations, Figs.(\ref{figIc}.a,c). Furthermore, the charge current maximum deviation Fig.(\ref{figIc}.b) occurs for disorder strength values ($U\geq 6$) larger than spin Hall maximum deviation ($U \approx 3$), Fig.(\ref{figIs}.b).
However, the spin Hall and charge currents deviations have the same behavior, the growth as a function of energy, Fig.(\ref{figIc}.d); For low energy ($E=0.02$) the charge current has its minimum deviation, while for high energies ($E \geq 0.6$) has its maximum. Hence, from the numeric data of Figs.(\ref{figIc}.b,d) we estimate the charge current maximum deviation as $\textbf{rms}[I^c] = e^2V/h \times 0.48$ (dashed line).

At this point, we can analyse the SHA, Eq.(\ref{angle}), which is depiceted in the Fig.(\ref{figASH}). The Figs.(\ref{figASH}.a,c) show the SHA average as a function of $U$ for different values of $\lambda$ and energy, respectively, while the Figs.(\ref{figASH}.b,d) are their respective deviations. As we can see in Figs.(\ref{figASH}.a,c), the SHA average  keeps the oscillations present in the spin Hall current average. However, the SHA maximum deviations happen only for $U\geq 6$ Fig.(\ref{figASH}.b), which means that the efficiency increase is not related with the spin Hall current fluctuations increase, but with the charge current fluctuations increase. The more the charge current fluctuates, the more efficient the charge-to-spin conversion, in accordance with Eq.(\ref{rmsT}).

\begin{figure}
\includegraphics[scale = 0.4]{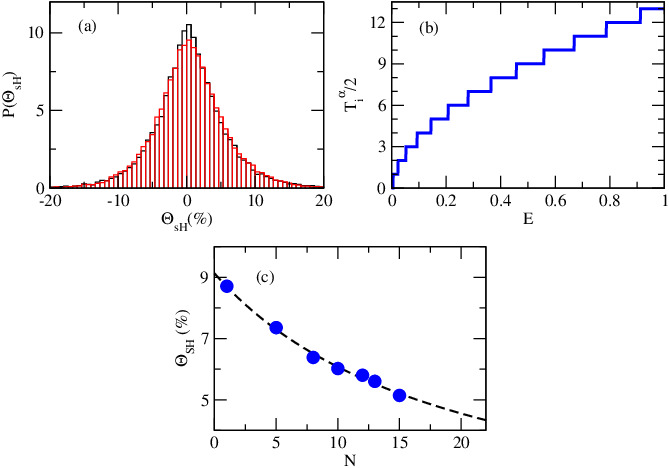}
\caption{(a) Histograms of SHA for $E=1$, $U=8$ and $\lambda =0.7, 0.8$. (b) The transmission coefficient $T_i^{\alpha}(E)/2=N$ as a function of energy. (c) The SHA maximum deviations of Fig.(\ref{figASH}.d) in function of thickness $N$. The dashed line is a numeric data fit.}\label{figAlNL}
\end{figure}

Although the Figs.(\ref{figIs}.d) and (\ref{figIc}.d) demonstrate an increase the maximum of the deviations with the energy, converging to a finite value, the SHA maximum deviations decrease with energy without the convergence, as demonstrated in the Fig.(\ref{figASH}.d). Therefore, for smaller energy $E=0.02$ the SHA has its maximum deviation $\Theta_{\text{sH}} \approx 9\% $, which means the SHA increasing with the energy decreasing, in agreement with Eq.(\ref{rmsT}). 

Finally, we are in a position to directly connect the numerical result and the CLT hypothesis/results, the Fig.(\ref{figASH}) and Eqs.(\ref{TM}) and (\ref{rmsT}). The  Fig.(\ref{figAlNL}) displays the connection. In the Fig.(\ref{figAlNL}.a) we plot the histograms of SHA for $E=1$, $U=8$ and $\lambda =0.7, 0.8$ and we demostrate the Gaussian distribution with zero average in accordance with CLT, as previously stated in Eq.(\ref{TM}). The Fig.(\ref{figAlNL}.b) shows the transmission coefficient $T_i^{\alpha}(E)=\sum_{j}\tau_{ij}^{\alpha}(E)=2N$ as a function of Fermi energy, which gives the relation between  $E=0.02, 0.2, 0.4, \dots$ and  $N=1,5, 8, \dots $. Hence, the Fig.(\ref{figAlNL}.c) shows the SHA maximum deviations of Fig.(\ref{figASH}.d) as a function of $N$. The dashed line is the numerical data fit, $\Theta_{\text{sH}} = (10.9+0.55 \times N)^{-1}$. Taking the limit of large values of $N$, for which the Eq.(\ref{rmsT}) is valid, it goes to  $\Theta_{\text{sH}} = 1.8 / N$. Comparing the latter with the Eq.(\ref{Tc}), we obtain $\sigma=Nl_e/L=N/ 10$, which drives to a universal relation $\Theta_{\text{sH}} \times \sigma = 0.18$, as previously stated.

{\it Conclusions} - In this work, we studied the SHA fluctuations of device in the disorderly regime with strong SOI. 
We were able to show that the SHA deviation depends only on three variables. Furthermore, in the limit which the sample length is long enough, the product between SHA maximum deviation and dimensionless conductivity holds a universal number, which is independent of material and its specific features. This universal relation is supported by an extensive theoretical numerical calculation. Beside, it was compared with five different experimental data showing in the Fig. (\ref{expprb}) obtained from Refs. [\onlinecite{PhysRevB.94.060412,doi:10.1021/acs.nanolett.6b02334,PhysRevB.98.094433,doi:10.1063/1.4753947,PhysRevB.101.094435}]. This result sheds light on the concept of SHE fluctuations and their importance in spintronic.

\begin{acknowledgments}
This work was supported by CNPq (Conselho Nacional de Desenvolvimento Cient\'{\i}fico e Tecnol\'ogico) and FACEPE (Funda\c{c}\~ao de Amparo \`a Ci\^encia e Tecnologia do Estado de Pernambuco).
\end{acknowledgments}

\end{document}